\documentclass[balance,upint,subscriptcorrection,varvw,mathalfa=cal=boondoxo,greek,pdf-a,colorlinks]{asmeconf}
\usepackage{stfloats}

\hypersetup{%
	pdfauthor={John H. Lienhard},
	pdftitle={ASME Conference Paper LaTeX Template},                  
	pdfkeywords={ASME conference paper, LaTeX template, BibTeX style},
	pdfsubject = {Describes the asmeconf LaTeX template},			  
	pdflicenseurl={https://ctan.org/pkg/asmeconf},
}

\begin{document}


\ConfName{Proceedings of the ASME 2023\linebreak International Mechanical Engineering Congress and Exposition}
\ConfAcronym{IMECE2023}
\ConfDate{October 29--November 2, 2023} 
\ConfCity{New Orleans, LA} 
\PaperNo{IMECE2023-XXXX}

\title{Improving CFD simulations by local machine-learned corrections}

 
%
%
%

\SetAuthors{%
	Peetak Mitra\affil{1}, 
	Majid Haghshenas\affil{1}, 
	Niccolo Dal Santo\affil{2}, 
	Conor Daly\affil{2},  
	David P.\ Schmidt \affil{1}\CorrespondingAuthor{schmidt@umass.edu}
	}

\SetAffiliation{1}{University of Massachusetts, Amherst, MA }
\SetAffiliation{2}{Mathworks, Milton, Cambridge}


\maketitle





\begin{abstract}

High-fidelity computational fluid dynamics (CFD) simulations for design space explorations can be exceedingly expensive due to the cost associated with resolving the finer scales. This computational cost/accuracy trade-off is a major challenge for modern CFD simulations. In the present study, we propose a method that uses a trained machine learning model that has learned to predict the discretization error as a function of large-scale flow features to inversely estimate the degree of lost information due to mesh coarsening. This information is then added back to the low-resolution solution during runtime, thereby enhancing the quality of the under-resolved coarse mesh simulation. The use of a coarser mesh produces a non-linear benefit in speed while the cost of inferring and correcting for the lost information has a linear cost.  We demonstrate the numerical stability of a problem of engineering interest, a 3D turbulent channel flow. In addition to this demonstration, we further show the potential for speedup without sacrificing solution accuracy using this method, thereby making the cost/accuracy trade-off of CFD more favorable.

\end{abstract}


\begin{nomenclature}
\EntryHeading{Roman letters}
\entry{$k$}{Turbulent kinetic energy}
\entry{$S$}{Shear rate tensor}
\entry{$\mathbf{S}$}{Source term for correcting the velocity solution}
\entry{$\mathbf{u}$}{Velocity}
\entry{$v$}{Cell volume}
\entry{$w$}{Weighting factor}

\EntryHeading{Greek letters}
\entry{$\Omega$}{Rotation rate tensor}
\entry{$\varepsilon$}{Error}
\entry{$\rho$}{Density}
\entry{$\delta$}{Mesh resolution}
\entry{$\mu$}{Fluid viscosity}
\entry{$\Phi$}{Solution field}
\entry{$\tau$}{Relaxation time}

\EntryHeading{Superscripts and subscripts}
\entry{$c$}{Coarse mesh value}
\entry{$f$}{Fine mesh value}
\entry{$\tau$}{Wall shear stress}
\entry{$v$}{Cell volume}
\end{nomenclature}


\section{Introduction}

Computational fluid dynamics (CFD) has become a cornerstone of modern engineering.  However, accurately predicting the large-scale features that usually drive the design process typically requires resolving small-scale features that are not as germane to the design process.  The necessary spatial and temporal resolution required to accurately model the physics and correctly predict the entire range of scales is often out of reach for many computational problems. While turbulence often garners much of the academic interest, the discretization error inherent in CFD is also of critical importance.  Turbulence can be modeled using Reynolds averaging (RANS) or Large Eddy Simulation (LES), but using coarse meshes for faster evaluations leads to the accumulation of discretization errors and therefore under-resolution of key features. This \textit{compute-accuracy} trade-off is a major driver of the cost of modern-day CFD.


In recent years improving or enhancing solution quality by using machine learning (ML), akin to image super-resolution, has become a major area of interest. The approaches range from using physics-constrained generative networks \cite{jiang2020meshfreeflownet} for full physics emulation, to building auto-differentiable frameworks that closely align the inductive biases of the ML algorithms to the physics \cite{shankar2020rapid, shankar2022validation} thereby aiding model interpretability and explainability. However, these cheap-to-investigate full physics surrogate methods suffer from the ability to generalize under unseen conditions as they lack explicit knowledge of the underlying governing equations. Kochov et al. \cite{kochkov2021machine} proposed using machine learning inside traditional fluid simulations, and suggested it can improve both the model accuracy and compute speed by an order of magnitude, and demonstrated the performance on canonical 2D examples. An alternate approach is to enhance solution quality of a under-resolved simulations by estimating the localized error. Coarsening the grid induces errors from primarily under-resolution as indicated by the modified partial differential equation \cite{celik2002discretization}. More recently, error surrogate models based on machine learning techniques have received much attention \cite{freno2019machine, hanna2020machine, drohmann2015romes}, largely because of their non-intrusive nature and fast on-line evaluations.  A review of several promising strategies by which machine learning can enhance CFD was published by Vinuesa and Brunton \cite{vinuesa2022}.

The principal contribution of this study is to make the cost-accuracy trade-off more favorable and demonstrate performance on an engineering-relevant 3D simulation. It is in the same vein that Kochkov et al. \cite{kochkov2021machine} demonstrated acceleration of LES simulations using ML based enhancement for the missing information in coarser meshes. Previous work in this area \cite{kochkov2021machine, watt2021correcting} showed the ML models have the ability to effectively super-resolve the missing information for applications ranging to 2D turbulence \cite{kochkov2021machine} and tracers in climate models \cite{brenowitz2020interpreting, watt2021correcting}. Several contributions have been made in error modeling for parameterized reduced-order models (ROM) \cite{drohmann2015romes, moosavi2018multivariate}, and the ideas have been extended to estimate discretization-induced errors \cite{freno2019machine}. Apart from some key differences in the implementation philosophy, a critical improvement over the previous work includes extending this approach to engineering relevant problems and to full 3D simulations. Our goal is to produce solutions to the Navier-Stokes equations with diminished sensitivity to mesh resolution.  In particular, we will focus on the velocity field since for the constant density Navier-Stokes, the velocity field and its derivatives sufficiently determine the pressure field. Therefore for a zero Mach number flow, such as in consideration here, the pressure field is neglected as part of the feature selection.

\section{Methods}

\begin{figure}
\centering
\includegraphics[width=0.4\textwidth]{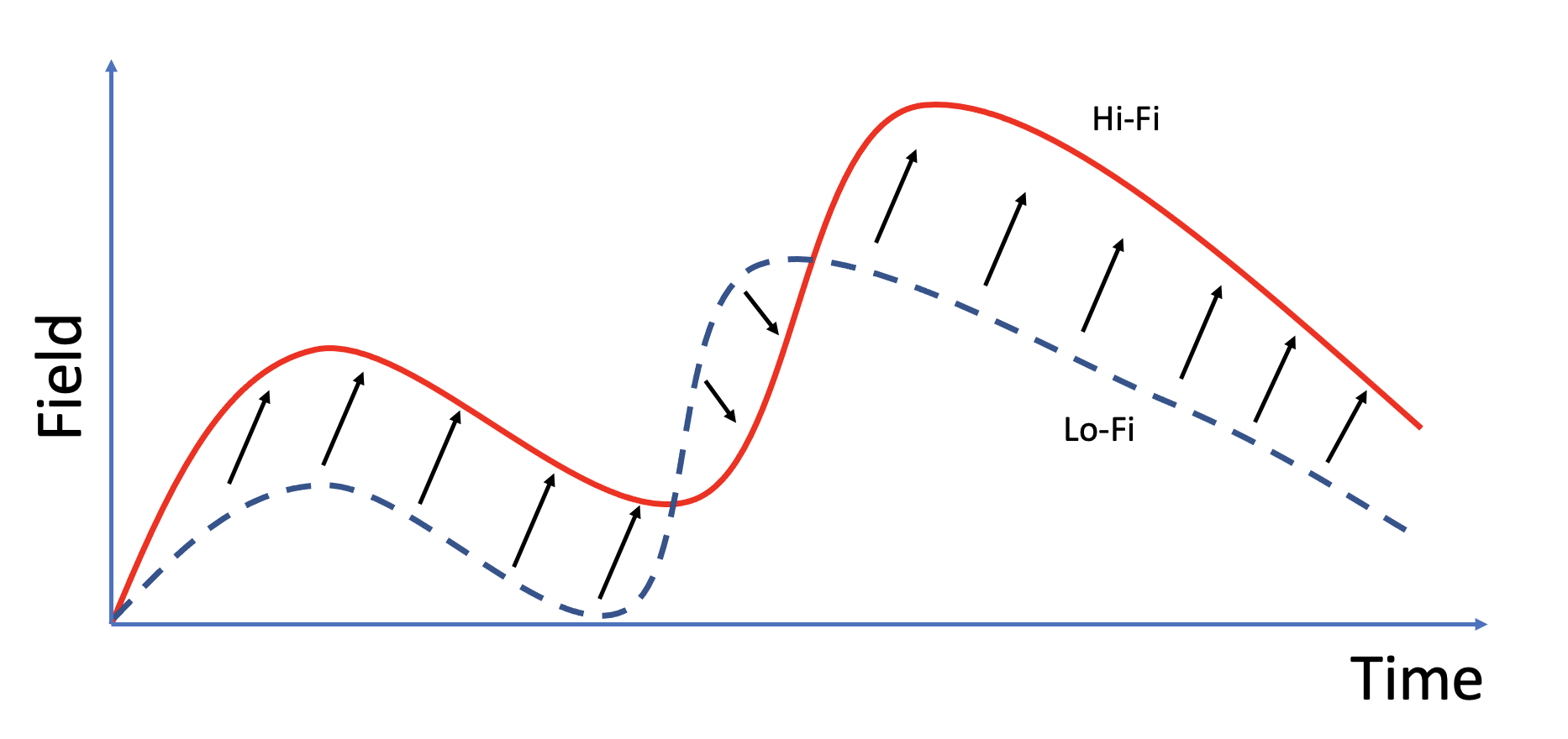}
\caption{\label{fig:Schematic} A schematic of the solution correction technique employed by the \textit{locally enhanced} approach, nudging the low-fidelity solution towards the more accurate solution\cite{watt2021correcting}. }
\end{figure}

The high-level functional premise of the \textit{local enhancement} method is shown in Figure \ref{fig:Schematic}. The proposed idea is to use corrections of local cell-level discretization error to nudge the lower fidelity (coarse grid) simulation towards the higher-fidelity solution. For a physical model system governed by a set of non-linear equations, the relationship between the high fidelity solution, $\mathrm{\Phi_{f}}$, from a fine mesh simulation and the coarse mesh predictions can be expressed as $ \mathrm{\Phi_f} = \mathrm{\Phi_{c}}( \mathrm{\delta}) + \mathrm{\varepsilon}$, where $\mathrm{\Phi_{c}}$ represents the solution field output of the low fidelity simulation from the coarse mesh with resolution $\delta$, and $\Phi_{f}$ represents the model variables - in our case fluid velocity, and $\mathrm{\varepsilon}$ the simulation error (lost information) due to numerical error. 

As explained above, for zero Mach number flow, the general field $\Phi$ is for this study specifically represented by $\mathbf{u}$, the fluid velocity. 
Functionally the error can be represented as, $\varepsilon = \mathbf{u_{f \rightarrow c}} - \mathbf{u_{c}}$, where subscripts $f$ and $c$ are fine and coarse respectively. The term $\mathbf{u_{f \rightarrow c}}$ is the fine to coarse mapped velocity, and $\mathbf{u_{c}}$  is the coarse mesh velocity. The additional step of mapping is an interpolation necessitated by the different node locations between a fine and a coarse mesh. Thus, to compute the local grid-induced error, it is necessary to map the fine-grid data $\Phi_{f}$ with resolution $\delta_f$ onto the coarse grid with resolution $\delta_c$. In other words, $\Phi_f$ is replaced by $\Phi_{f\rightarrow c}$ which is the fine-grid field of $\Phi$ mapped on a grid whose cell length is $\delta_c$.  This mapping, or interpolation, constitutes a source of error as some details of the flow field profile are lost due to interpolation. Using higher-order interpolation techniques, we minimize this source of additional error to $\mathcal{O}(10^{-5})$. This is achieved by using OpenFOAM's \cite{jasak2007openfoam} in-built \textit{mapFields} functionality. The locally enhanced velocity within each cell would then have the functional form, $\mathbf{u_{e}} = \mathbf{u_{c}} + \mathrm{LC}(\mathbf{u_{c}} )$, where $\mathbf{u_{e}}$ is the enhanced velocity, $\mathbf{u_{c}}$ is the coarse grid velocity, and $\mathrm{LC}$ is the learned correction provided by the machine learning algorithm, during inference time. The basic assumption for the application of the coarse-grained approach is that the coarse mesh simulation is able to capture/resolve the basic flow features. It would be inconceivable to use ultra-coarse representations of the physics such that any important detail is not resolved by the coarse mesh, thereby extrapolating the mapping abilities for the machine learning algorithm.

\subsection{Machine learning algorithm}

The inductive biases of the data are a point-to-point correlation. For example, the error $\mathrm{\varepsilon}$, is based on the cell-level information lost between the mapped and the low-resolution solution. Since there are no spatial or temporal correlations to be learned and in aligning with the inductive biases of the problem itself, we make use of a deep feed-forward neural network as our machine learning algorithm. Functionally, the machine learning model $f$ is learning the relationship between the coarse mesh (input) to the error (target),

\begin{equation}
    \mathrm{\varepsilon} = f(\mathrm{\Phi_{c}})
\end{equation}

The model training procedure involves the following steps:

\begin{itemize}
    \item \textbf{Run a fine-mesh simulation}: This simulation typically consists of a very
large number of cells and therefore is very accurate. 
\item \textbf{Run many coarse mesh simulations}: Run simulations with different coarse mesh configurations.
This step explores the multi-dimensional space in which error is created and provides input data for training.
\item \textbf{Mapping fine mesh solution to coarse mesh stencil}: Use OpenFOAM’s \cite{jasak2007openfoam} in-built mapping functionalities to map the fine
data generated in Step 1 onto the coarser stencil from Step 2. This would be
our ground truth that the model aspires to achieve. In machine learning jargon, this
is called the target data.
\end{itemize}

Similar to tuning constants in a physics model, hyperparameters such as network width, depth, and learning rate in deep neural networks represent the largest source of uncertainity in model outputs. This study involved conducting a Bayesian optimization based shallow neural architecture search \cite{mitra2020effectiveness} to identify the strongest candidates for the key hyperparameters. In the end the deep network was trained using 8 layers and 48 neurons in each layer. The initial learning rate was set to 0.0002 with a cosine learning rate decay \cite{loshchilov2016sgdr}. The optimizer used in this study was Adam \cite{kingma2014adam}.

Many engineering-relevant CFD simulations are inherently transient and often times this leads to the presence of outliers (tails in a distribution) in input data. This is especially true for scenarios involving moving geometries as well as complex and intermittent physics such as combustion. While in machine learning there are different best practices to deal with such outlier data, they often indicate ignoring them as it leads to poor training and generalization abilities for the model. Using this framework for building a surrogate model might lead to a loss of important, transient physics and therefore lead to the degrading performance of the model itself. One method to alleviate this problem is to use a customized loss function. Compared to the mean-squared-error or L2 loss, which amplifies the outliers, a mean-absolute-error or L1 loss, tends to fit the mean better. Our proposed loss makes use of both of these losses in a weighted fashion. The weighting between the losses is based on the data distribution of outliers. The training loss used in this study functionally can be represented as

\begin{equation}
    \mathit{loss} = w * L2 + (1-w) * L1
\end{equation}

where $w$ is set to 0.7 for the current study.

Once trained, for run-time inference, we freeze the deep network graph and convert it into its C++ equivalent that is compatible to use with the OpenFoam library. The full details of integrating a trained neural network to OpenFoam's C++ library have been discussed in previous studies \cite{haghshenas2020turbulence, mitra2021numerical}.

\subsection{Modified Governing Equations}

The machine-learned correction, or "nudge," is integrated into the Navier-Stokes governing equations by adding a source term, $\mathbf{S}$, represented by $\mathbf{u_{f \rightarrow \Delta}} - \mathbf{u_c}$. The modified governing equations functionally are shown as below.

\begin{equation}
    \frac{\partial \mathbf{u}}{\partial t} + \mathbf{u} \cdot \triangledown \mathbf{u} = - \frac{\triangledown P}{\rho} +  \triangledown \nu_{\mathit{eff}} \triangledown\mathbf{u} + \frac{\mathbf{S}}{\tau}
\end{equation}

where $\mathbf{u}$ is the fluid velocity vector, $P$ is the fluid pressure, $\rho$ is the fluid density, $\nu_{\mathit{eff}}$ is the effective kinematic viscosity.  The effective viscosity is the sum of the molecular and turbulent contributions.  The term $\tau$ is used as an arbitrary time-scale factor. In other words, it is used to relax the amount of extra information (machine-learned nudge) that is added to the system of governing equations. This is primarily done to ensure numerical stability for the non-linear PDE solution. Add too much source, and the mass conservation has a hard time keeping the solution stable and converging. Add too little source, and the solution barely changes. This is therefore a hyperparameter in the modeling setup and we empirically investigate the effects of different relaxation factors.  A more scientific intuition or explanation is therefore warranted, and is a subject of future work. The explicit treatment is expected to be less stable numerically and subject to von Neumann stability considerations. However, the stability can be considerably improved by appropriately choosing robust solvers such as the Preconditioned Bi-Conjugate Gradient.

\subsection{Problem setup}

Our test case is a 3D turbulent channel flow with turbulent Reynolds number ($\mathrm{Re_{\tau}}$) of 395 simulated with LES. For full details of the geometry and details, interested readers are referred to the original DNS study \cite{moser1999direct}. To simulate an infinite domain, periodic boundary conditions are commonly applied in the stream- and spanwise directions. The pressure gradient is then introduced via an extra forcing term in the momentum equations. The turbulent channel flow is a statistically-developing internal flow through parallel smooth walls. The x-axis is the mean flow direction, the y-axis is the wall normal and the z-axis is the spanwise direction with statistically homogeneous flow with periodic boundaries.  Therefore most of the contribution to the velocity field is in the x-direction, with diminishing contributions from the other two axes. This means that a neural network trained on the x-component of velocity (or the error therein) can be used as a surrogate for the entire velocity magnitude, due to the large contributions. The turbulence model chosen in generating the fine mesh and coarse mesh data is the LES wall adapting local eddy viscosity model (WALE) model \cite{nicoud1999subgrid}. 

This case will allow using a mesh fine enough to resolve the larger turbulent structures present in the flow, yet small enough for the case to be computed in a reasonable time on a single workstation. The greatest challenge in the LES simulation is that in the near-wall region of a turbulent boundary layer the necessary resolution required of a high-quality LES renders such simulations expensive unless a high degree of empiricism is introduced into the sub-grid modeling process \cite{duraisamy2019turbulence}. Then the challenge for the machine learning algorithm is to not only learn the bulk flow error but also the near wall error accurately.

\subsection{Inputs}

The Helmholtz decomposition theorem states that every smooth vector field $\mathbf{u}$ can be decomposed into a rotational part and an irrotational part \cite{sommerfeld2016mechanics}. Hence, the motion of a fluid element can be defined in terms of three fundamental components (pure translation motion, pure strain along the principal axes, and rotation rate) \cite{sommerfeld2016mechanics}. Therefore in choosing our inputs we prioritize using the strain rate tensor and the rotation rate tensor. This is particularly useful to preserve Galilean and rotational invariance, to prevent any directional preferences that the model may learn. We appropriately non-dimensionalize tensors, as well as use non-dimensional quantities such as cell Reynolds number and wall distance as inputs. The non-dimensionalization factors were inspired from \cite{wu2018physics} and are tabulated in Table \ref{wrap-tab:0} where $R_{c}$ is the cell Reynolds number, measured as $Re \equiv \frac{\rho u \delta}{\mu}$ where $\delta$ is the cube root of the cell volume, $S$ is the shear-rate tensor, $\Omega$ is the rotation-rate tensor and $\mathbf{Y}$ is the wall distance. The non-dimensionalization factors for each term are discussed in Table \ref{wrap-tab:0}.

\begin{table}
\centering
\caption{Input Features and its Normalization factors}\label{wrap-tab:0}
\begin{tabular}{cc}\\\toprule  
 Input & Normalization Factor \\\midrule
 $\triangledown \mathbf{u}$ & $\frac{\sqrt{k}}{\delta_V}$ \\  \midrule
  $S$ & $\frac{\sqrt{k}}{\delta_V}$ \\ \midrule
$\Omega$ & $\frac{\sqrt{k}}{\delta_V}$ \\  \midrule
$$Re$$ & -- \\  \midrule
$\mathbf{Y}$ & $\delta_V$ \\ \bottomrule
\end{tabular}
\end{table}

\subsection{Quantitative metrics}

In addition to qualitative metrics to measure performance we define a quantitative criterion to measure success for the locally enhanced approach. It is defined as the \textit{cell volume weighted L2 norm} defined as $\mathrm{L2} = \sum \delta * (\mathbf{u_{f \rightarrow c}} - \mathbf{u})^2$ where $\delta$ is the cube root of cell volume, $\mathbf{u_{f \rightarrow c}}$ is the ground truth velocity mapped to the coarser CFD mesh, and $\mathbf{u}$ is the velocity predicted by CFD (locally enhanced or coarse-mesh simulation). We choose to focus on the velocity error as it is the metric we use to locally enhance the coarse mesh simulation. A lower L2 norm of error would establish the improvement in accuracy of the enhanced result compared to the coarse mesh simulation. 

The use of coarse-graining reduces the cell count in the mesh.  This reduction is quantified by a mesh Reduction Factor (RF), defined as the number of cells in the fine-grained mesh used to produce that ground truth data set divided by the number of cells in the coarse mesh.  Because the cell size connects to the cost per iteration of the linear solvers, the number of iterations required per time step, and the time step size, the relationship between the reduction factor and overall computational cost is expected to be non-linear.  With a larger RF, the opportunity for the learned correction to accelerate the computation and reduce the error is greater.

\section{Results}

The training data are sampled across three different turbulent Reynolds numbers of 290, 395 and 500. Further, the simulations at each Reynolds number consist of fourteen different coarse mesh configurations. The reduction factor, defined as the ratio of fine mesh cells to coarse mesh cells, ranged from 1.12 to 4.5 for a total fine mesh cell count of 60,000. The training dataset comprised of about 8 million points. The \textit{a priori} performance on a test data yields a $R^2$ of 0.8460, which indicates a reasonable fit. The large range of learning, in terms of the near wall behavior, mean flow characteristics across different discretizations and large scale flow configurations, are some of the challenges to achieving a precise regression.

\begin{figure*}[h]
    \centering
    \includegraphics[width=0.95\linewidth]{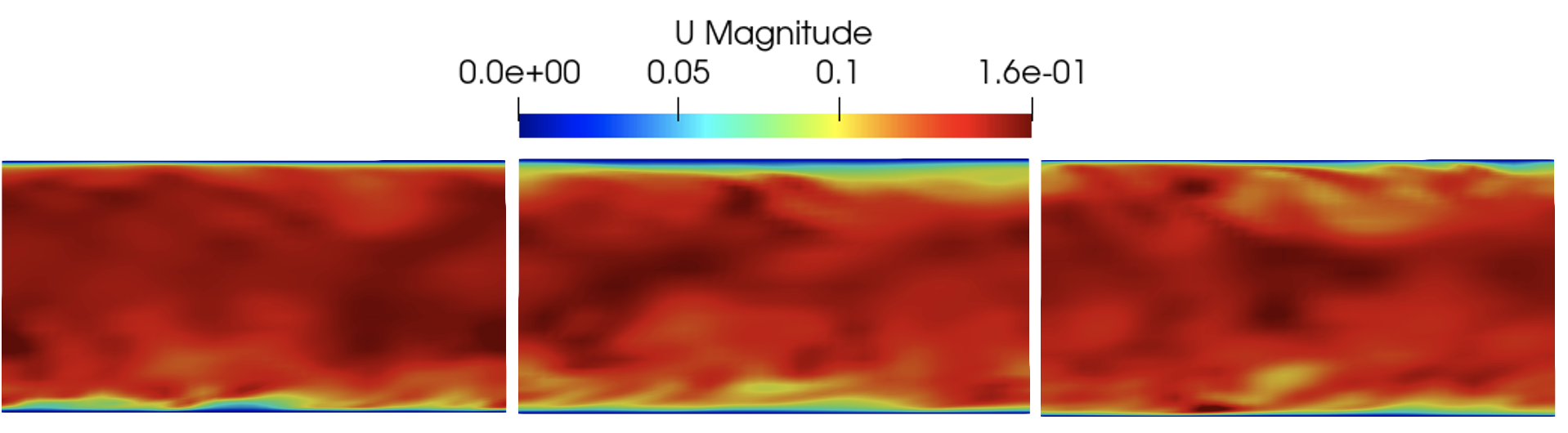}
    \caption{The midplane clips suggest the locally enhanced simulation (right panel) is able to recover lost near wall information, thereby lowering errors and improving time to solution.  The panel at the left is from the fine to coarse mesh mapping, the
middle panel is from the coarse mesh simulation, and the right panel is from the
network-enhanced simulation. Each plan shows the mid-clip plane colored by the
velocity magnitude (scaled similarity). The network enhanced
(right panel) recovers missing information (ground truth in the left panel) compared
to the coarse mesh (middle panel) simulation.}
    \label{fig:Umag}
\end{figure*}

The trained network is coupled to the OpenFOAM \cite{jasak2007openfoam} solver \textit{pimpleFoam}.  The qualitative performance for the velocity magnitude is indicated by examination of a mid-clip plane in Figure \ref{fig:Umag}. This snapshot is taken at time t=1000s, for a mesh reduction factor of 2. The simulation was run on a fine mesh, considered to be the ground truth, which is then interpolated to the coarse mesh (left panel) for comparison to coarse-mesh CFD.  The results of CFD run on the coarse mesh (middle panel) fail to accurately resolve the near-wall effects seen in the left panel. On the other hand, the error-corrected results from coarse-mesh simulation (right most) recover a large degree of lost information near the walls. 

\begin{figure*}[h!]
\centering
\includegraphics[width=0.95\textwidth]{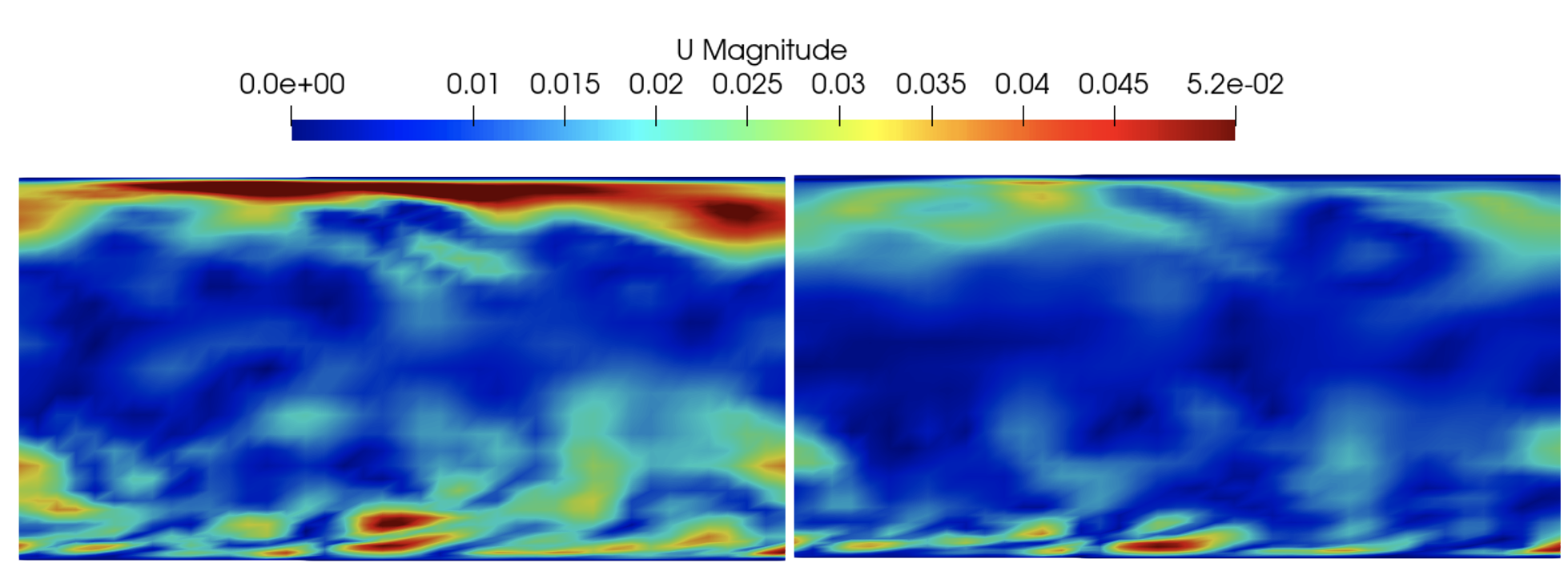}
\caption{\label{fig:Udiff}  The velocity magnitude difference between the
mapped (ground truth) and the CFD simulations.  The left panel shows the uncorrected coarse-mesh discrepancy. The locally enhanced simulation (right panel) is able to recover lost information in the near-wall region thereby improving solution accuracy. The differences in the velocity magnitude further confirm the earlier observation that the network enhanced (right panel) recovers missing information,
and therefore has lower velocity magnitude differences.}
\end{figure*}

\begin{figure*}
    \centering
    \includegraphics[width=0.9\linewidth]{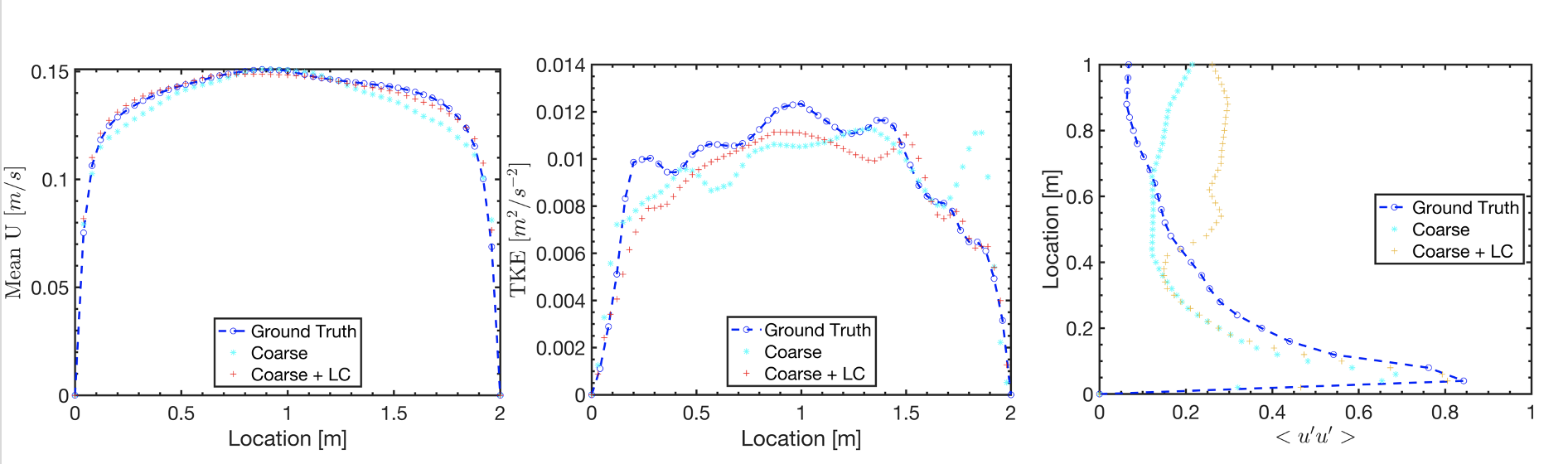}
    \caption{Each panel shows plots of  three different curves. One for the mapped field, one from the coarse mesh simulations, and one from the coarse mesh enhanced simulation. The left most panel shows the time-averaged behavior and indicates information recovery for the coarse mesh enhanced simulation and a consistent tracking of the mapped field (ground truth data). The middle and the right panel are from instantaneous Turbulent Kinetic Energy and Reynolds stress behavior in the near wall region. It is evident that the enhanced simulation recovers near wall behavior better compared to the low resolution coarse mesh simulation. }
    \label{fig:VelPlot}
\end{figure*}

Figure \ref{fig:Udiff} presents the velocity magnitude difference between the mapped (ground truth) and the CFD simulations. The left panel shows the difference between mapped and coarse mesh simulation, and the observation in the near wall region behavior is consistent with the earlier result. The right panel is the difference (shown on the same scale) between the mapped and the locally enhanced coarse mesh simulation. It is clearly evident that the locally enhanced simulation is able to recover lost information, especially close to the wall boundary. The time-averaged x-direction velocity performance is reported in Figure \ref{fig:VelPlot}. The vertical line probes are placed at the center of the channel at 2m from the channel entrance (total length of the channel is 4 m). Figure \ref{fig:VelPlot} shows the time-averaged x-component of velocity at the 2m location and it is evident that the locally enhanced simulations improve the solution performance and recover lost information, especially in the near-wall region and in the mean flow. The middle and the right panel in Figure \ref{fig:VelPlot} represents the instantaneous snapshots of the Turbulent Kinetic Energy and the Reynolds stress tensor (in the near wall region), showing the degree of improvement in the prediction for the learned correction model. For the turbulent channel flow, the near wall region is the most challenging to resolve and very important from the perspective of viscous dissipation, and energy generation.

In addition to the qualitative diagnostics, the $\mathrm{L2}$-norm of the error is calculated for the entire range of reduction factors. The error data are shown in Table \ref{wrap-tab:1}. In comparing the coarse and the locally enhanced simulation performance for each reduction factor, it is observed that the additional source term improves the simulation fidelity significantly, up to an order of magnitude. The largest gain is obtained for the higher degree of coarseness. This is understandable since for a very coarse mesh, the loss of details is proportionately higher and therefore the learned correction model has a larger impact in the accuracy recovery.

\begin{figure*}
\centering

\includegraphics[width=0.95\textwidth]{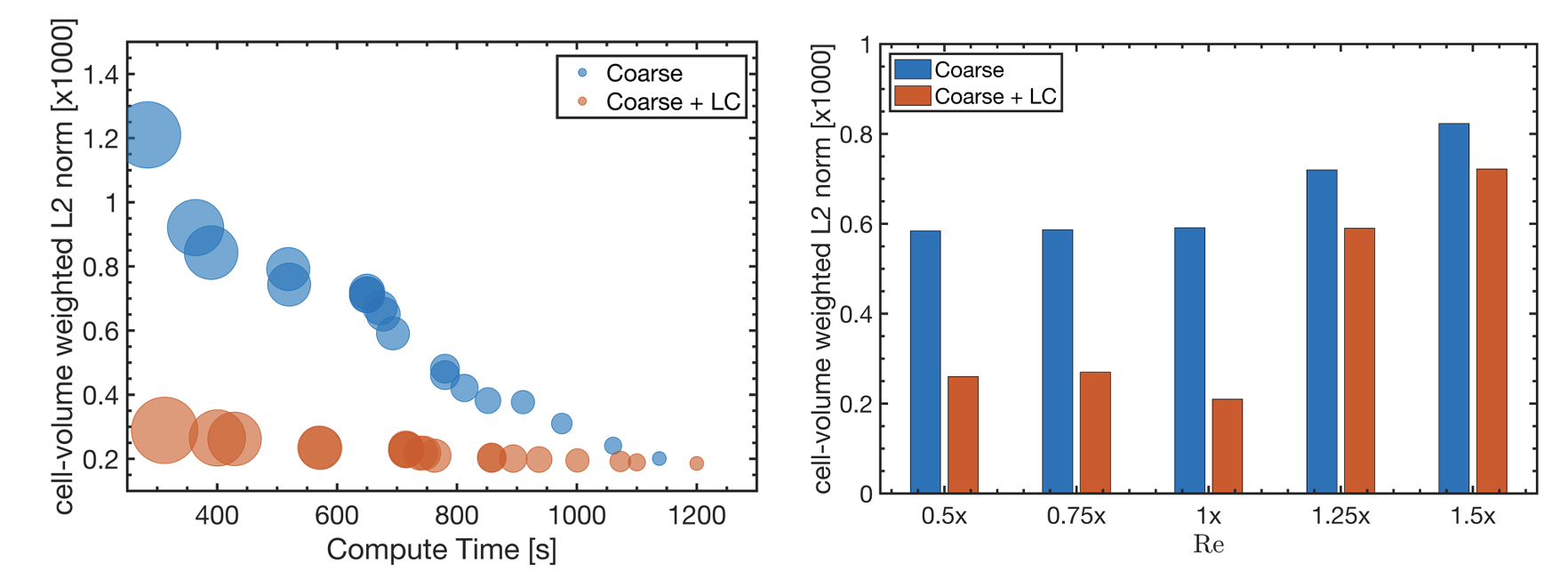}
\caption{\label{fig:L2} Both panels indicate the local enhancement is able to provide better solutions at a lower cost, even for unseen run conditions. The timing plot on the left shows the relative improvement in the cost versus
accuracy, as a result of the local enhancement.  The size of the circles indicates the reduction factor of the cell count.  The right panel shows the norm of the error at a range of Reynolds numbers, indicating the ability of the scheme to work at other Reynolds numbers. }
\end{figure*}

The left panel in Figure \ref{fig:L2} represents the compute cost versus accuracy trade-off for the turbulent channel flow study. The radius of each of the circles represents the reduction factor. The larger the reduction factor (or coarser the mesh), the larger the radius of the circle. Comparing circles of similar sizes gives a measure of the performance gains with the local enhancement approach. The general trend is that by using the local enhancement approach, there is a potential for massive gains in reducing errors (and therefore improving solution accuracy), for a moderate increase in solution cost (for example, ML-enhanced solutions add about 10\% on average cost to the time to solution). One other way to look at this is to compare the coarse mesh circles (blue) with the local enhanced circles (orange) along the Y-axis. To obtain an error norm of 0.35, the coarse mesh simulations took about 1000s (wall time), whereas similar levels of accuracy were obtained at a fraction of the compute cost in approximately 300s, thereby indicating a compute speed-up of over 3x for a similar fidelity solution. The speedup can be further improved by studying larger problems, which are expected to be more expensive to compute. This increased cost would result in higher information retrieval at a fraction of the cost making the cost-accuracy trade off even more beneficial.  Whereas the computational cost of CFD increases non-linearly with the cell count, the cost of the learned correction is linear.  Kochkov et al. \cite{kochkov2021machine} used a 2D DNS dataset for their ground truth and reported 40-80x speed ups. The cost to perform DNS on this channel flow is orders of magnitude higher than the fine mesh LES employed here and therefore there are performance gains yet to be realized using this locally enhanced approach.

\begin{table}
\centering
\caption{Performance Improvements from the Learned Correction.  Reduced resolution, quantified as a reduction factor,  is listed versus the percent reduction in error.}\label{wrap-tab:1}
\begin{tabular}{cc}\\\toprule  
 Mesh reduction Factor & \% Error reduction \\\midrule
4.57 & 76.11\\  \midrule
3.33 & 70.25\\  \midrule
2.50 & 68.51\\  \midrule
2.00 & 67.31 \\ \midrule
1.52 & 48.14 \\ \midrule
1.14 & 7.67 \\ \bottomrule
\end{tabular}
\end{table}

\section{Conclusions}

A machine learning mesh error correction algorithm has been developed and implemented within open-source 3D CFD code OpenFOAM. This error correction allows a CFD simulation to achieve higher fidelity with lower resolution.  The numerical stability of this method is demonstrated on a full 3D CFD simulation, relevant to many engineering applications.   The approach achieved 3-5x speedups with minimal reduction in observed accuracy. An advantage of the locally enhanced method is its mesh invariance. For example, some of the current approaches for solution enhancement are limited by using Cartesian mesh, whereas there are no such requirements for this locally enhanced approach. The artificial time-scale term ($\tau$) is a hyperparameter and is currently chosen empirically for this study. A more scientifically rigorous method of choosing it is desirable and a subject of future work.  Further ways to extract more performance benefits that could be realized by attacking larger problems, where the linear cost of the algorithm would generate additional speedup.

\bibliographystyle{unsrt}
\bibliography{bib.bib}

\end{document}